\documentclass[superscriptaddress,twocolumn,floats,showpacs,prl,amsmath,amssymb,floatfix,nofootinbib,balancelastpage]{revtex4}

\input epsf
\usepackage{psfig}

\newcommand{\mnras}{Mon. Not. R. Astr. Soc.}

\newcommand{\wjm}{\left(
                           \begin{array}{ccc}
         l_1 & l_2  & l_3  \\
         m_1 & m_2  & m_3
                           \end{array}
                   \right)}

\newcommand{\Ylm}[1]{Y_{l_#1}^{m_#1}}

\newcommand{\Ylmn}{Y_{l}^{m}}

\newcommand{\alm}[1]{a_{l_#1 m_#1}}

\newcommand{\bea}{\begin{eqnarray}}
\newcommand{\eea}{\end{eqnarray}}
\newcommand{\bean}{\begin{eqnarray*}}
\newcommand{\eean}{\end{eqnarray*}}

\newcommand{\bx}{{\bf x}}

\newcommand{\rad}{{r}}

\newcommand{\bn}{{\bf \hat{n}}}

\newcommand{\bm}{{\bf \hat{m}}}

\begin{document}

\title{21-cm Background Anisotropies Can Discern Primordial Non-Gaussianity}
\author{Asantha Cooray}
\email{acooray@uci.edu}
\affiliation{Department of Physics and Astronomy, 4186 Frederick Reines Hall, University of California, Irvine, CA 92697}

\begin{abstract}
The non-Gaussianity of initial perturbations provides information on the mechanism that generated
primordial density fluctuations.   We find that 21-cm background anisotropies due to
inhomogeneous neutral hydrogen distribution prior to reionization captures
information on primordial non-Gaussianity better than a high-resolution cosmic microwave background anisotropy
map. An all-sky 21-cm experiment over the frequency range from 14 MHz to 40 MHz 
with angular information out to a multipole of 10$^5$
can limit the primordial non-Gaussianity parameter $f_{\rm NL} \lesssim 0.01$. 
\end{abstract}

\pacs{98.70.Vc,98.65.Dx,95.85.Sz,98.80.Cq,98.80.Es}

\maketitle

\noindent \emph{Introduction--- } The cosmic 21-cm background involving spin-flip line emission or absorption  
of neutral hydrogen contains unique signatures  on how the neutral gas evolved from last 
scattering at $z \sim 1100$ to complete reionization at $z \sim 10$ \cite{Furlanetto}.
Subsequent to recombination, the temperature of neutral gas is coupled to that of the CMB. At redshifts below
$\sim$ 200 the gas cools adiabatically, its temperature drops below that of the CMB, and neutral hydrogen
resonantly absorbs CMB flux through the spin-flip transition \cite{field,loeb}. The inhomogeneous
neutral hydrogen density distribution generates anisotropies in the brightness temperature measured 
relative to the blackbody CMB \cite{zaldarriaga}. The large cosmological
information content in 21-cm background is well understood in the
literature \cite{loeb,Pen,sigurdson}. If the primordial fluctuations
are non-Gaussian, then the 21-cm anisotropies will naturally contain a signature associated with that  non-Gaussianity.

In this {\it Letter}, we show that the angular bispectrum of 21-cm anisotropies contains a measurable non-Gaussianity from primordial
fluctuations, if the scale-independent quadratic corrections to the gravitational potential captured by $f_{\rm NL}$ has a value 
above $\sim$ 0.01. This is substantial given that even a perfect CMB experiment limited by cosmic variance alone can
only restrict $f_{\rm NL} > 3$ \cite{Komatsu} while there is no significant improvement when  
using low redshift large-scale structure \cite{Scoc}.
In comparison, the expected non-Gaussianity under standard inflation is of order $|n_s-1|$,
and with the scalar spectral index $n_s \sim 0.98$ \cite{Spergel}, 
this non-Gaussianity is well below unity \cite{Maldacena}.
The primordial non-Gaussianity parameter $f_{\rm NL}$, however, has a correction
associated with evolution of second and higher-order perturbations after inflation \cite{Bartolo}.
For standard inflation $f_{\rm NL} = -5/12(n_s-1) + 5/6 +3/10f(k)$ where the last term is
momentum dependent. In this case, a primordial non-Gaussianity is always present and  $f_{\rm NL}$ is at the level of a few tenths.

Thus, compared to all other probes of inflationary parameters,
such as the spatial curvature and amplitudes and spectral indices of scalar and tensor spectra, 
$f_{\rm NL}$ is the only parameter for which we 
do not yet have an observational probe to reach the simple expectation. 
Here, we establish 21-cm background as a potentially useful  probe to study $f_{\rm NL}$ values below one.
When compared to CMB, 21-cm background has two distinct advantages: 
(1) the ability to probe multiple redshifts based on frequency selection
and (2) the lack of a damping tail in the 21-cm anisotropy  
spectrum, similar to early damping in CMB spectrum at a multipole around 2000.

Here, we also consider the confusion from other non-Gaussian signals in the 21-cm sky.
We introduce the two-to-one correlator statistic \cite{Coo01} 
that can be optimized to detect the primordial bispectrum by appropriately filtering 21-cm anisotropy data. This statistic has already been
used to limit $f_{\rm NL}$ with CMB data \cite{Szapudi}.  
The statistic is optimized with a
filter to extract most information on the amplitude of the
primordial non-Gaussianity, but
requires prior knowledge on the configuration dependence
 of the primordial 21-cm bispectrum.
The amplitude of non-Gaussianity remains a free parameter to be determined
from the data. The {\it Letter} is organized as following: w
e first discuss the bispectrum in 21-cm anisotropies associated with
primordial perturbations resulting from quadratic corrections to the primordial potential.
We discuss ways to measure this bispectrum in the presence of other non-Gaussian signals and
determine the extent to which $f_{\rm NL}$ can be measured from 21-cm background data.

\begin{figure}[!t]
\centerline{\psfig{file=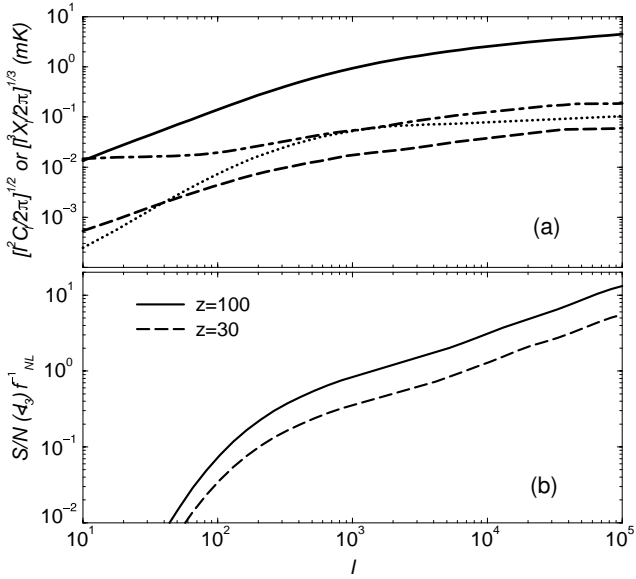,width=3.3in,angle=0}}
\caption{(a) The 21-cm anisotropy power spectrum ($\sqrt{l^2C_l/2\pi}$; solid line) 
and  non-Gaussian squared 21-cm anisotropy---21-cm anisotropy power spectrum ($[l^3X_l/2\pi]^{1/3}$; see text for details)
with the optimal filter applied (see, equation~12). The dashed line is the signal related  to the primordial 
bispectrum with $f_{\rm NL}=1$, while the dotted line is the noise bias $X_l^{\rm grav}$   from
 perturbation theory bispectrum under non-linear gravitational clustering.
The dot-dashed line shows the total noise spectrum, $[l^3N_l/2\pi]^{1/3}$, associated with the measurement of $X_l^{\rm prim}$.
These spectra are calculated at $z=100$. (b) The cumulative signal-to-noise ratio for a detection of
the non-Gaussian spectrum $X_l$ as a function of the multipole  $l$ at $z=100$ (solid line) and 
at $z=30$ (dashed line). These estimates assume no detector noise with measurements limited by cosmic variance alone and with $f_{\rm sky}=1$.}
\label{cl}
\end{figure}

\noindent \emph{21-cm Signal--- } The 21-cm anisotropies are observed as a change in the intensity of the
CMB due to line emission or absorption at an observed frequency $\nu$:
\begin{equation}
T_b(\bn,\nu) =  \frac{T_S - T_{\rm CMB}}{1+z} \, \tau(\bn,\nu)
\label{eq:dtb}
\end{equation}
where $T_S$ is the spin temperature of the neutral gas, $z$ is the
redshift corresponding to the frequency of observation ($1+z=\nu_{21}/\nu$, with
$\nu_{21} = 1420$ MHz) and $T_{\rm CMB} = 2.73 (1+z) K$ is the CMB temperature at redshift $z$. 
The optical depth, $\tau$, in the hyperfine transition \cite{field}, when
accounted for density and velocity perturbations of a patch in the neutral hydrogen distribution \cite{loeb,zaldarriaga,Bharadwaj}, is
\begin{eqnarray}
\tau & = & \frac{ 3 c^3 \hbar A_{10} \, \bar{n}_{\rm H}}{16 k_B \nu_{21}^2 \, T_S \, H(z) } 
\left(1+\delta_H - \frac{1+z}{H(z)}\frac{\partial v}{\partial r}\right)  \, ,
\label{eq:tauigm} 
\end{eqnarray}
where $A_{10}$ is the spontaneous emission coefficient for the transition ($2.85 \times 10^{-15}$ s$^{-1}$),
$n_{\rm HI}$ is the neutral hydrogen density, 
$\delta_H=(n_{\rm H}-\bar{n}_{\rm H})/\bar{n}_{\rm H}$ is the inhomogeneity in the density,
$v$ is the peculiar velocity of the neutral gas distribution, $r$ is the comoving 
radial distance, and $H(z)$ is the expansion rate at a redshift of $z$.  For simplicity, we have dropped the dependences
in location $\bn$. The fluctuations in the CMB brightness temperature is
\begin{equation}
\delta T_b=\bar{T}_b\left[\left(1-\frac{T_{\rm CMB}}{\bar{T}_S}\right)\left(\delta_H- \frac{1+z}{H(z)}\frac{\partial v}{\partial r}\right) + \frac{T_{\rm CMB}}{\bar{T}_S} S \delta_H\right]  \, ,
\label{eqn:deltab}
\end{equation}
where $S(z)$ describes the coupling between fluctuations in the spin temperature and the neutral density distribution \cite{Bharadwaj}.
In above $\bar{T}_b \approx 26.7 {\rm mK} \sqrt{1+z}$.  

In terms of non-Gaussian measurements, we discuss the angular averaged bispectrum of 21-cm anisotropies 
constructed with spherical harmonic moments of the 21-cm background defined as
\begin{equation}
a_{l m}(\nu) = \int d\bn Y^*_{l m}(\bn) \delta T_b(\bn,\nu)\, .
\end{equation}
For simplicity in notation, we will drop the explicit dependence on the frequency, but it should
be assumed that 21-cm background measurements can be constructed as a function of frequency,  and thus as a function of
redshift, with the width in frequency space primarily limited by the bandwidth of a radio interferometer.

To generate a non-Gaussianity in the 21-cm background, we assume quadratic corrections to the
primordial potential fluctuations such that $\Phi(\bx)=\Phi_L(\bx)+f_{\rm NL} (\Phi^2_L(\bx)-\langle \Phi^2_L(\bx)\rangle)$ when $\Phi_L(\bx)$ is the linear and Gaussian  perturbation and $f_{\rm NL}$ is the coupling constant, which may
or may not be scale dependent. Under this description, existing WMAP data limits $-30 < f_{\rm NL} < 74$ 
at the 1$\sigma$ confidence \cite{Szapudi}, 
while with an ideal CMB experiment fundamentally limited by cosmic variance  one can constrain $|f_{\rm NL}| < 3$ \cite{Komatsu}. As discussed, this is well above the standard expectations for $f_{\rm NL}$ 
even after accounting for second-order evolution during horizon exit and re-entry \cite{Bartolo}.

Using the resulting three-dimensional bispectrum for potential fluctuations, and noting that 21-cm observations trace the density field,
we can write the expected angular bispectrum of 21-cm background as
\begin{eqnarray}
&& B^{\rm prim}_{l_1 l_2 l_3} = \nonumber \\
&& \sum_{m_1 m_2 m_3} \wjm \left[
\left< a_{l_1 m_1}a_{l_2 m_2}a_{l_3 m_3}  \right> + {\rm Perm.} \right]\nonumber \\
&=& A_{l_1,l_2,l_3} \left(\frac{16}{\pi^3}\right) \int_0^\infty \rad^2 d\rad \, 
\left[ b_{l_1}(\rad) b_{l_2}(\rad) b^{\rm NL}_{l_3}(\rad) + {\rm Perm.} \right]\, .
\label{eqn:ovbidefn}
\end{eqnarray}
Here,
\begin{eqnarray}
&&A_{l_1,l_2,l_3}=\sqrt{\frac{(2l_1 +1)(2 l_2+1)(2l_3+1)}{4 \pi}}
\left(
\begin{array}{ccc}
l_1 & l_2 & l_3 \\
0 & 0  &  0
\end{array}
\right) \nonumber \\
&& b_{l}(\rad) = \int_0^\infty d\rad' \int_0^\infty \frac{k^2 dk}{M(k)}  P_{\delta \delta}(k,\rad') 
f_k(\rad') W_\nu(\rad') j_l(k\rad) \nonumber \\
&& b^{\rm NL}_{l}(\rad) = \int_0^\infty d\rad' \int_0^\infty \frac{k^2 dk} {M(k)^{-1}} \frac{f_{\rm NL}}{D(\rad')}
f_k(\rad') W_\nu(\rad') j_l(k\rad) \, . \nonumber \\
\label{eqn:bl}
\end{eqnarray}
In above,
\begin{equation}
f_k(\rad) = \bar{T}_b(\rad) \left[\left(1-\frac{T_{\rm CMB}}{\bar{T}_S}\right)J_l(k\rad) + \frac{T_{\rm CMB}}{\bar{T}_S}Sj_l(k\rad)\right] \, ,
\end{equation}
with $J_l(x)=j_l(x)-j_l''(x)$.
In equation~(\ref{eqn:bl}), $D(\rad)$ is the growth function of density perturbations, $M(k)=-3k^2T(k)/5\Omega_mH_0^2$ maps the 
primordial potential fluctuations to that of the density field, 
and $W_\nu(\rad)$ is the window function of observations related to the central frequency and the bandwidth.
In equation~(\ref{eqn:ovbidefn}) permutations account for a total of three terms with the replacement of $l_2 \rightarrow l_3$ and $l_1$.
For reference, the angular power spectrum of 21-cm anisotropies follows from above as 
\begin{equation}
C_l(\nu) = \frac{2}{\pi} \int k^2 dk P_{\delta \delta}(k) \left[\int_0^\infty d\rad D(\rad) W_\nu(\rad) f_k(\rad)\right]^2 \,.
\end{equation}

While we have explicitly written out the bispectrum related to primordial non-Gaussianity, 21-cm anisotropies will contain
additional signatures of non-Gaussianity. Among these is the 21-cm bispectrum
resulting from the non-linear gravitational evolution of 
density perturbations with a three-dimensional bispectrum of the form $B_{\delta}(k_1,k_2,k_3)=F({\bf k_1},{\bf k_2})P_{\delta \delta}(k_1)P_{\delta \delta}(k_2)+{\rm Perm.}$, where $F({\bf k_1},{\bf k_2})$ is the mode-coupling kernel \cite{Fry}. 
We denote the resulting angular bispectrum of 21-cm anisotropies sourced by the non-linear density field as $B^{\rm grav}_{l_1,l_2,l_3}$ when the above three dimensional bispectrum is substituted to the
formalism outline with equation~(5) \cite{Lig}.
Moreover, once the universe begins to reionize, 21-cm anisotropies are modulated by fluctuations in the neutral fraction, $x_{\rm H}$,
 in addition to density  perturbations with $\tau \propto \bar{n}_{\rm H}\bar{x}_{\rm H}(1+\delta_{\rm H})(1+\delta_{\rm x})$.
Though small in amplitude, the second-order corrections to the brightness temperature lead to a non-Gaussian signal \cite{Coo05}.
As we are concentrating on redshifts prior to reionization, this secondary bispectrum can be safely ignored.

For a measurement of $f_{\rm NL}$ we consider a simple statistic that captures non-Gaussian information
once an expected configuration for the 21-cm bispectrum is specified.
The statistic involves the power spectrum constructed with the filtered 2-to-1 correlator \cite{Coo01,Szapudi}:
\begin{eqnarray}
X(\bn,\bm) &\equiv& \langle \hat{T}_b^2(\bn) T_b(\bm) \rangle  \\
           &=& \sum_{l_1 m_1 l_2 m_2} \langle \alm{1}^{2} \alm{2}^* 
\rangle
               \Ylm{1}(\bn) \Ylm{2} {}^*(\bm)\, , \nonumber
\label{eqn:twopointsquared}
\end{eqnarray}
where $a_{lm}^2 = \int d\bn \hat{T}_b^2(\bn) \Ylmn {}^*(\bn)$, where $\hat{T}_b(\bn)$ is the 21-cm brightness temperature with the filter 
described below applied in multipole space.
The measurable statistic can be written as 
\begin{eqnarray}
\langle a_{lm}^2 a^*_{l'm'} \rangle &=&  \left[X_l^{\rm prim}+X_l^{\rm grav}\right]\delta_{ll'}\delta{mm'} \nonumber \\
X_l^{\rm (i)} &=& \sum_{l_1 l_2} B^{\rm (i)}_{l_1 l_2 l} w_{l_1,l_2|l}  \frac{A_{l,l_1,l_2}}{(2l+1)} \, ,
\label{eqn:finalform}
\end{eqnarray}
where $(i)$ represents the 21-cm bispectrum generated by either primordial non-Gaussianities or non-linear gravitational evolution.
Since the bispectrum is defined by a triangle in multipole space
with lengths of sides $(l,l_1,l_2)$, the $X_l$ statistic essentially captures information
from all triangular configurations of the bispectrum with one of the sides fixed at length $l$.  
The filter function $w(l_1,l_2|l)$ 
in multipole space, however, is applied so that one can
maximize the information related to one form of non-Gaussianity
with a specified configuration dependence. Here, this allows a mechanism to
separate the primordial non-Gaussianity from other non-Gaussian signals
since we have prior theoretical expectation on the
the expected configuration dependence  \cite{Bartolo}. 

The signal-to-noise ratio for detecting $X_l^{\rm prim}$ with an experiment covering
 $f_{\rm sky}$ fraction of the sky    is ${\rm S}/{\rm N} = \sqrt{f_{\rm sky} \sum_l (X_l^{\rm prim})^2/(N_l^{\rm tot})^2}$, where
$N_{l}^{\rm tot}$ is the total noise contribution which involves both the noise bias from
$X_l^{\rm grav}$ and the Gaussian variance in the proposed statistic:
\begin{eqnarray}
&&N_l^{\rm tot}  = \left[\frac{\left(X_l^{\rm grav}\right)^2}{2l+1} + 
\frac{C_l^{\rm tot}}{(2l+1)^2} 
\sum_{l_1 l_2} C_{l_1}^{\rm tot} C_{l_2}^{\rm tot} w_{l_1,l_2|l}^2 A_{l,l_1,l_2}^2\right]^{1/2} \, .
\label{eqn:noise}
\end{eqnarray}
Here,  $C_{l}^{\rm tot}$ represents all contributions to the 21-cm anisotropy power spectrum such that
$C_{l}^{\rm tot} = C_l^{\rm 21-cm} + C_l^{\rm noise}$
when $C_l^{\rm noise}$ is the instrumental noise contribution.
Note that we are treating the confusion, $X_l^{\rm grav}$, in the same manner foreground signals 
are treated in CMB studies. 

As suggested above, to maximize the signal-to-noise ratio for a detection of $X_l^{\rm prim}$,
following Ref.~\cite{Coo01} we found the filter to be of the form
\begin{equation}
w^{\rm prim}_{l_1,l_2|l} = \frac{b^{\rm prim}_{l,l_1,l_2}}{C_{l_1}^{\rm tot}C_{l_2}^{\rm tot}} \, ,
\end{equation}
where $b^{\rm prim}_{l,l_1,l_2} = B^{\rm prim}_{l,l_1,l_2}/A_{l,l_1,l_2}$.  The filter is normalized such that $\sum_{l_1 l_2} w^{\rm prim}_{l_1,l_2|l} =1$.
The filter is written in terms of $B^{\rm prim}_{l,l_1,l_2}$, but due to normalization of the filter, is independent of $f_{\rm NL}$, thus
no prior knowledge on the amplitude of the non-Gaussianity is required. What is necessary, however, is
the expected configuration dependence, which can be 
specified a priori under an assumed theoretical model for primordial non-Gaussianity.

Since the filter prescribes the expected bispectrum configuration,
it also acts as a way to reduce the amplitude of
confusions. For example, with the filter 
applied, $X_l^{\rm grav} \propto \sum_{l_1l_2} B_{l,l_1,l_2}^{\rm grav} B_{l,l_1,l_2}^{\rm prim}$
and  as modes do not align exactly, there is a significant cancellation. We find that the resulting
noise spectrum for a measurement of  $X_l^{\rm prim}$ is dominated by the Gaussian variance captured by the second term of
equation~(\ref{eqn:noise}) and not the confusion resulting from $X_l^{\rm grav}$. 
This statement is independent of $f_{\rm NL}$.
If stated differently, when properly filtered to search for the primordial non-Gaussianity,
the main confusion for detecting primordial signal is not the non-Gaussianity generated by
non-linear perturbations  but rather the Gaussian covariance associated with 
the statistical measurement of $X_l^{\rm prim}$.  

Finally, while we  have not explored here as the Gaussian variance dominates, it may be possible to
further optimize the filter such that the additional non-Gaussian confusions cancel out exactly. 
In practice, the measurement of $X_l^{\rm prim}$ is likely to be further confused by the non-Gaussianity of 
foregrounds.
Unfortunately, little is known about the expected level of the foreground intensity in the frequency range of interest. 
Techniques have been suggested and discussed to remove foregrounds below the detector noise levels \cite{Morales,Santos} and the filtering process we have outlined will further reduce the confusion from the remaining residual foregrounds. This is clearly a topic for further study
 once data become available with first-generation interferometers\footnote{http://www.lofar.org; 
http://www.haystack.mit.edu/arrays/MWA}.

\noindent \emph{Results--- }  Fig.~1(a) summarizes the power-spectrum of 21-cm anisotropies generated by the neutral hydrogen distribution at a redshift of 100. Here, we also plot $X_l^{\rm prim}$, $X_l^{\rm grav}$,
and $N_l$ for the same redshift with the optimal filter applied. As shown, $N_l > X_l^{\rm grav}$, 
suggesting that the noise term is dominated by the Gaussian variance (second term in equation~11). 
Note that we have estimated $N_l$ under the 
assumption that observations are limited only by the cosmic variance.  This allows us to establish the potentially achievable
limit and compare directly with cosmic variance limit of $f_{\rm NL}$ with CMB.

In Fig.1(b), we summarize the estimate related to signal-to-noise ratio for a detection of $X_l^{\rm prim}$ as a function  of $l$.
The typical signal-to-noise ratio, when measurements are out to a multipole of $10^5$, is at the level of $\sim$ $13 f_{\rm NL}$,
with observations centered at a redshift of 100, to $\sim 5 f_{\rm NL}$, at a redshift of 30.
If one restricts the calculation to certain configurations, such as equilateral triangles of
the bispectrum, then the cumulative signal-to-noise ratio is $\sim 0.3  f_{\rm NL}$ \cite{pillepich}.
Out to a multipole of $10^5$, however, there is a total of $\sim 10^{15}$ modes in the whole bispectrum,
whereas with a specific configuration, one is only restricting to at most 10$^5$ modes.
Our statistic $X_l$ captures all this information and is optimally filtered so as to maximize information
on the primordial non-Gaussianity. When using all the information, 
though there are  10$^{15}$ modes, the filtering selection  removes 
a large number of modes when separating the primordial signal from 
the secondary non-Gaussian signal.

Furthermore, 21-cm observations lead to measurements at multiple redshifts.
One cannot, however, make arbitrarily small bandwidths to improve the detection since at scales below a 
few Mpc, anisotropies in one redshift bin will be correlated with those in adjacent bins \cite{Santos}.
With a bandwidth of 1 MHz, corresponding to a radial distance of 30 Mpc at a redshift of $\sim 50$,
over the frequency interval  between 14 MHz ($z \sim 100$) to 45 MHz ($z\sim 30$),
we find a cumulative signal-to-noise ratio, ${\rm S}/{\rm N} = \sqrt{\sum_z [{\rm S}/{\rm N}(z)]^2}$, 
of $\sim 10^2 f_{\rm NL} \sqrt{f_{\rm sky}}$ out to $l_{\rm max}$ of 10$^5$. In return,
one can potentially probe $f_{\rm NL}$ values as low as $10^{-2}$. 
With first-generation  instruments, we would at most survey 1\% of the sky. 
Assuming instrument noise is dominating at multipoles above $10^4$ between 30 MHz and 60 MHz at 1 MHz
bandwidths, we find a signal-to-noise ratio of $0.5 f_{\rm NL} \sqrt{f_{\rm sky}/0.01}$,
which is equivalent to the ultimate limit from CMB.

The lowest $f_{\rm NL}$ value one can reach with 21 cm data is
two orders of magnitude better than CMB \cite{Komatsu}.
This is the main result of the paper with important implications for future cosmological studies. 
While it may be challenging for first-generation interferometers, low frequency 21-cm observations should
eventually establish the primordial non-Gaussianity. The limitation to achieve this detection
is not a fundamental one,  such as cosmic variance,
but rather that involving adequate technology for high-sensitive low-frequency radio observations. 
In the same manner  technology has been steadily improving over the last two decades 
to allow  high precision  CMB measurements, it is very likely that experiments will
progress rapidly allowing precise 21-cm studies. While we lacked  a unique probe to reach $f_{\rm NL}$ 
values below 0.1, our calculations suggest that this is no longer the case  once we open the low-frequency 
radio sky for observations.

\end{document}